\documentclass[11pt]{article}
\usepackage{amsfonts}

\usepackage{amssymb}
\usepackage{latexsym}
\usepackage{graphicx}
\usepackage[english]{babel}

\begin{document}
\input epsf

\makeatletter
\@addtoreset{equation}{section}
\makeatother


\begin{flushright}
\end{flushright}
\vspace{20mm}

 \begin{center}
{\LARGE Deforming the D1D5 CFT away from the orbifold point}
\\
\vspace{18mm}
{\bf   Steven G. Avery$^\Diamond$\footnote{avery@mps.ohio-state.edu}, Borun D. Chowdhury$^\sharp$\footnote{b.d.chowdhury@uva.nl} and Samir D. Mathur$^\Diamond$\footnote{mathur@mps.ohio-state.edu}
\\}
\vspace{8mm}
$\Diamond$ 
Department of Physics,\\ The Ohio State University,\\ Columbus,
OH 43210, USA\\ 
\vspace{8mm}
$\sharp$
Instituut voor Theoretische Fysica,\\
Universiteit van Amsterdam,\\
Amsterdam 1018 XE, The Netherlands\\
\end{center}
\vspace{10mm}

\thispagestyle{empty}
\begin{abstract}

The D1D5 brane bound state is believed to have an `orbifold point' in its moduli space which is the analogue of the free Yang Mills theory for the D3 brane bound state. The supergravity geometry generated by D1 and D5 branes is described by a different point in moduli space, and in moving towards this point we have to deform the CFT by a marginal operator: the `twist' which links together two copies of the CFT.  In this paper we find the effect of this deformation operator on the simplest physical state of the CFT -- the Ramond vacuum. The twist deformation leads to a final state that is populated by pairs of excitations like those in a squeezed state.  We find the coefficients characterizing the distribution of these particle pairs (for both bosons and fermions) and thus write this final state in closed form.

\end{abstract}
\newpage
\renewcommand{\theequation}{\arabic{section}.\arabic{equation}}

\def\p{\partial}
\def\h{{1\over 2}}
\def\be{\begin{equation}}
\def\bea{\begin{eqnarray}}
\def\ee{\end{equation}}
\def\eea{\end{eqnarray}}
\def\r{\rightarrow}
\def\tildr{\tilde}
\def\n{\nonumber}
\def\nn{\nonumber \\}
\def\t{\tilde}
\def\b{\bigskip}
\newcommand{\Nsc}{\mathcal{N}}
\newcommand{\bj}{\bar{\jmath}}
\def\sqi{{1\over \sqrt{2}}}
\newcommand\eqref[1]{(\ref{#1})}

\newcommand\ket[1]{|#1\rangle}
\newcommand\bra[1]{\langle #1|}
\newcommand\com[2]{[#1,\,#2]}
\newcommand\ac[2]{\{#1,\,#2\}}

\section{Introduction}
\label{intr}\setcounter{equation}{0}

The D1D5 bound state is perhaps the best system to tackle the physics
of black holes. This system gives a nonzero entropy at extremality,
both for the 2-charge D1D5 bound state and its excitation, the
3-charge D1D5P bound state~\cite{counting}.
Non-extremal excitations of this state collide and exit the bound
state at exactly the rate at which Hawking radiation is produced from
the corresponding black hole~\cite{radiation}. The microstate
structure of 2-charge states and a large number of 3-charge states has
been found, and the `fuzzball' nature of these states resolves the
well known Hawking information paradox since these microstates do not
have a traditional horizon~\cite{fuzzballs}.

AdS/CFT duality \cite{adscft} relates the D1D5 CFT to
the gravity solution produced by the bound state. But the CFT has been
mostly studied at its `free' point, which is believed to be an
`orbifold CFT' \cite{orbifold}. The CFT at this point is dual
to a gravitational solution which is very singular, and not in a
domain interesting for gravitational physics like the formation of
black holes. To get a description of interesting gravitational
phenomena we will have to move in the moduli space of the CFT to a
point away from the `orbifold' point. In particular we need to turn on
the `blow up' mode of the orbifold, which is given by a `twist
operator' in the orbifold CFT.  Such a deformation has been discussed
in various contexts in earlier work~\cite{deformation}.

\begin{figure}[ht]
\begin{center}
\includegraphics[width=4cm]{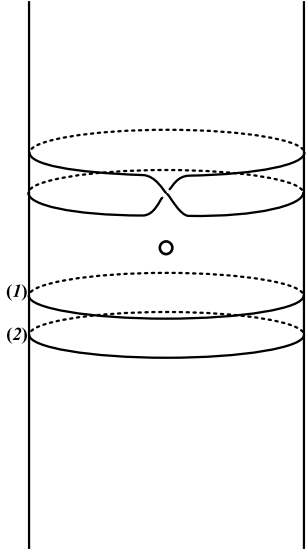}
\caption{The effect of the twist contained in the deformation operator: two circles at earlier times get joined into one circle after the insertion of the twist.}\label{2-twist}
\end{center}
\end{figure}

Our goal in this paper will be to study the effect of this deformation on excitations of the CFT. The orbifold CFT is given by a 1+1 dimensional sigma model with target space $({\cal M}_4)^{N_1N_5}/S_{N_1N_5}$, the symmetric product of $N_1N_5$ copies of a 4-manifold ${\cal M}$. Here ${\cal M}$ can be $T^4$ or $K3$; we will take it to be $T^4$. We will take the spatial circle $\sigma$ of the CFT to be compact. Each copy of $T^4$ gives rise to a free $c=6$ CFT. Fig.\ref{2-twist} shows the effect of the twist operator: it takes two copies of the $c=6$ CFT  and links them together to make one copy of the CFT living on a doubly wound circle. 

In this paper we do the simplest computation involving the deformation operator. We start with the vacuum state for each of the two initial copies of the $c=6$ CFT. In the physical problem of the black hole the vacuum states are in the Ramond sector, and we will let both copies have the `spin down' Ramond ground state. We then act with the deformation operator at a location $w_0=\tau_0+i\sigma_0$ on the cylinder describing the 1+1 dimensional CFT. For $\tau>\tau_0$ we have one copy of the $c=6$ CFT living on a doubly wound circle, with a set of bosonic and fermionic excitations that are created by the effect of the deformation operator. Our goal is to find the state at $\tau>\tau_0$. We argue that the excitations in this state have the structure of excitations in a squeezed state (schematically $\sim e^{\gamma a^\dagger a^\dagger}$ for the bosons, and similarly for the fermions).  Thus we find the state by finding the coefficients $\gamma_{mn}$ for the bosonic and fermionic excitations. The deformation operator also has  a supersymmetry current $G$ applied to this squeezed state, and after taking that into account we write down the full final state obtained by the action of the deformation operator on our chosen vacuum state. 

To use this result for the physics of the D1D5 system we should integrate over the location $w_0$, and also allow different possible excitations in the initial state. We will carry out those steps elsewhere, and restrict ourselves to finding the basic squeezed state here.

\section{The D1D5 CFT at the orbifold point}

\subsection{The CFT}

Consider type IIB string theory, compactified as
\be
M_{9,1}\rightarrow M_{4,1}\times S^1\times T^4.
\label{compact}
\ee
Wrap $N_1$ D1 branes on $S^1$, and $N_5$ D5 branes on $S^1\times
T^4$. The bound state of these branes is described by a field
theory. We think of the $S^1$ as being large compared to the $T^4$, so
that at low energies we look for excitations only in the direction
$S^1$.  This low energy limit gives a conformal field theory (CFT) on
the circle $S^1$.

We can vary the moduli of string theory (the string coupling $g$, the
shape and size of the torus, the values of flat connections for gauge
fields etc.). These changes move us to different points in the moduli
space of the CFT. It has been conjectured that we can move to a point
called the `orbifold point' where the CFT is particularly simple
\cite{orbifold}. At this orbifold point the CFT is
a 1+1 dimensional sigma model. The 1+1 dimensional base space is
spanned by $(y, t)$, where
\be
0\le y<2\pi R
\ee
is a coordinate along the $S^1$, and $t$ is the time of the 10-d
string theory. For our CFT computations, we rotate time to Euclidean
time, and also use scaled coordinates $(\sigma, \tau)$ where the space
direction of the CFT has length $2\pi$:
\begin{equation}
\tau = i\frac{t}{R} \qquad \sigma = \frac{y}{R}.
\end{equation}
We continue back to Lorentzian signature at the end.

The target space of the sigma model is the `symmetrized product' of
$N_1N_5$ copies of $T^4$,
\be
(T_4)^{N_1N_5}/S_{N_1N_5},
\ee
with each copy of $T^4$ giving 4 bosonic excitations $X^1, X^2, X^3,
X^4$. It also gives 4 fermionic excitations, which we call $\psi^1,
\psi^2, \psi^3, \psi^4$ for the left movers, and $\bar\psi^1,
\bar\psi^2,\bar\psi^3,\bar\psi^4$ for the right movers. The fermions can be
antiperiodic or periodic around the $\sigma$ circle. If they are
antiperiodic on the $S^1$ we are in the Neveu-Schwarz (NS) sector, and
if they are periodic on the $S^1$ we are in the Ramond (R)
sector\footnote{The periodicities flip when we map the cylinder to the complex
plane because of a Jacobian factor.}. The central charge of the theory with fields
$X^i, \psi^i, ~i=1\dots 4$ is
\be
c=6
\ee
The total central charge of the entire system is thus $6 N_1N_5$.

\subsubsection{Symmetries of the CFT}

The D1D5 CFT has $(4,4)$ supersymmetry, which means that we have
$\mathcal{N}=4$ supersymmetry in both the left and right moving
sectors. This leads to a superconformal ${\cal N}=4$ symmetry in both
the left and right sectors, generated by operators $L_{n}, G^\pm_{r},
J^a_n$ for the left movers and $\bar L_{n}, \bar G^\pm_{r}, \bar
J^a_n$ for the right movers. The algebra generators and their OPEs and commutators are given in Appendix~\ref{ap:CFT-notation}. 

Each $\Nsc = 4$ algebra has an internal R symmetry group
SU(2),\footnote{In fact, the full R symmetry group of the
$\mathcal{N}=4$ algebra is $SO(4)$; however, the other $SU(2)$ does
not have a current associated with it within the algebra.} so there is
a global symmetry group $SU(2)_L\times SU(2)_R$.  We denote the
quantum numbers in these two $SU(2)$ groups as
\be
SU(2)_L: ~(j, m);~~~~~~~SU(2)_R: ~ (\bj, \bar m).
\ee
In the geometrical setting of the CFT, this symmetry arises from the
rotational symmetry in the 4 space directions of $M_{4,1}$ in
Equation~\eqref{compact},
\be
SO(4)_E\simeq SU(2)_L\times SU(2)_R.
\label{pthree}
\ee
Here the subscript $E$ stands for `external', which denotes that these
rotations are in the noncompact directions. These quantum numbers
therefore give the angular momentum of quanta in the gravity
description.  We have another $SO(4)$ symmetry in the four directions
of the $T^4$. This symmetry we call $SO(4)_I$ (where $I$ stands for
`internal'). This symmetry is broken by the compactification of the
torus, but at the orbifold point it still provides a useful organizing
principle. We write
\be
SO(4)_I\simeq SU(2)_1\times SU(2)_2.
\ee
We use spinor indices $\alpha, \dot\alpha$ for $SU(2)_L$ and $SU(2)_R$
respectively. We use spinor indices $A, \dot A$ for $SU(2)_1$ and
$SU(2)_2$ respectively.

The 4 real fermions of the left sector can be grouped into complex
fermions $\psi^{\alpha A}$ with the reality constraint
\be
 (\psi^\dagger)_{\alpha A}=-\epsilon_{\alpha\beta}\epsilon_{AB} \psi^{\beta B}
\ee
The right fermions have indices $\bar{\psi}^{\dot\alpha \dot A}$ with
a similar reality constraint. The bosons $X^i$ are a vector in the
$T^4$. They have no charge under $SU(2)_L$ or $SU(2)_R$ and are
given by
\begin{equation}
[X]_{\dot{A}A} = {1\over \sqrt{2}}X^i(\sigma^i)_{\dot{A}A}.
\end{equation}
where $\sigma^i, i=1, \dots 4$ are the three Pauli matrices and the identity. (The
notations described here are explained in full detail in
Appendix~\ref{ap:CFT-notation}.)

\subsection{States and operators}

Since we orbifold by the symmetric group $S_{N_1N_5}$, we generate
`twist sectors', which can be obtained by acting with `twist
operators' $\sigma_n$ on an untwisted state. Suppose we insert a
twist operator at a point $w_0$. As we circle the
point $w_0$, different copies of $T^4$ get mapped into each other. Let
us denote the copy number by a subscript $a=1, 2, \dots n$. The twist
operator is labeled by the permutation it generates. For instance,
every time one circles the twist operator
\be
\sigma_{(123\dots n)},
\label{qone}
\ee 
the fields $X_i^{(a)}$ get mapped as
\begin{equation}
X_i^{(1)} \rightarrow
X_i^{(2)} \rightarrow
\cdots
\rightarrow
X_i^{(n)} \rightarrow X_i^{(1)},
\end{equation}
and the other copies of $X_i^{(a)}$ are unchanged. We have a similar
action on the fermionic fields. Each set of linked copies of the CFT is called
one `component string'.

The simplest states of the CFT are in the `untwisted sector' where no copy of the $c=6$ CFT is linked to any other copy; i.e. all component strings have winding number unity.  Consider one component string, and consider the theory defined on the cylinder.  The fermions on this string can be either periodic around the $\sigma$ circle of the cylinder (Ramond sector R) or antiperiodic (Neveu-Schwarz sector NS). Consider one copy of the $c=6$ CFT. The simplest state of this theory is the NS sector vacuum 
\be
|0\rangle_{NS}: ~~~h=0, ~~m=0
\label{nsvac}
\ee
But the  CFT arising from the D1D5 brane bound state is in the Ramond (R)
sector. One can understand this because the periodicities of the
fermions around the $S^1$ are inherited from the behavior of fermionic
supergravity fields around the $S^1$ in \eqref{compact}. These
supergravity fields must be taken to be periodic, since otherwise we
would generate a nonzero vacuum energy in our spacetime and the metric far from the branes would not be flat. We can relate the state 
(\ref{nsvac}) to a Ramond ground state using spectral flow \cite{spectral}. Spectral flow maps amplitudes in the CFT to amplitudes in another CFT; under this map  dimensions and charges change as (we write only the left sector)
\be
h'=h+\alpha j +{c\alpha^2\over 24}, ~~~
j'=j+{c\alpha\over 12}
\label{spectral}
\ee
We have $c=6$. Setting $\alpha=-1$ gives
\be
|0^-_R\rangle: ~~h={1\over 4}, ~~~m=-\h
\ee
which is one of the Ramond ground states of the $c=6$ CFT for a component string with winding number unity. Other Ramond ground states are obtained by acting with fermion zero modes, so that we have four states in all
\be
|0_R^-\rangle,~~~\psi^{++}_0|0_R^-\rangle, ~~~\psi^{+-}_0|0_R^-\rangle, ~~~\psi^{++}_0\psi^{+-}_0|0_R^-\rangle
\label{rground}
\ee
(with similar possibilities for the right moving sector).

The deformation operator involves the twist $\sigma_2$. As we go around a point of insertion of this twist, the fermions in the first copy change to fermions in the second copy, and after another circle return to their original value. Creating such a twist automatically brings in a `spin field' at the insertion point, which has $h={1\over 2}, j=\h$ \cite{lm2}. Thus there are two possible insertions of such a twist, with $m=\h$ and with $m=-\h$. We write these as
$\sigma_2^+$ and $\sigma_2^-$
respectively. The operator $\sigma_2^+$ is a chiral primary and $\sigma_2^-$ is an anti-chiral primary. 

\subsection{The deformation operator}

Let us describe the deformation operator in some detail.

\subsubsection{The structure of the operator}

The deformation operator is a singlet under $SU(2)_L\times SU(2)_R$.
To obtain such a singlet we apply modes of $G^\mp_{\dot A}$ to
$\sigma_2^\pm$. A singlet under $SU(2)_L\times SU(2)_R$ can be made as
(writing both left and right moving sectors)~\cite{deformation} 
\be
\hat O_{\dot A\dot B}\propto{1\over 4}\epsilon_{\alpha\beta}\epsilon_{\dot\alpha\dot \beta}  \Big [{1\over 2\pi i}\int_{w_0} dw G^\alpha_{\dot A}(w)\Big ]\Big [{1\over 2\pi i} \int_{\bar w_0} d\bar w {\bar G}^{\dot\alpha}_{\dot B}(\bar w)\Big ]\sigma_2^{\beta\dot\beta}\label{deformation}
\ee
In Appendix~\ref{ap:EqOfTwist} we show that
\be
{1\over 2\pi i} \int _{w_0} dw G^-_{\dot A} (w) \sigma_2^+(w_0)\propto {1\over 2\pi i} \int _{w_0} dw G^+_{\dot A} (w) \sigma_2^-(w_0)
\ee
Thus we can write the deformation operator as (we chose its normalization at this stage)
\be
\hat O_{\dot A\dot B}(w_0)=\Big [{1\over 2\pi i} \int _{w_0} dw G^-_{\dot A} (w)\Big ]\Big [{1\over 2\pi i} \int _{\bar w_0} d\bar w \bar G^-_{\dot B} (\bar w)\Big ]\sigma_2^{++}(w_0)
\ee
The normalization of $\sigma_2^{++}$ will be specified below. The indices $\dot A, \dot B$ indices can be contracted  to rewrite the above four operators as a singlet and a triplet of $SU(2)_1$.\footnote{Since we can write the deformation operator in terms of $G^-\sigma_2^+$ or in terms of $G^+\sigma_2^-$, it provides a good check on the results that we get the same final state each way.}

\subsubsection{Normalization of $\sigma_2^+$}\label{normsection}

Now we describe how we normalize $\sigma_2^{++}$. Let us focus on the left moving part of the operator, which we denote by $\sigma_2^+$. Let the conjugate operator be called $\sigma_{2, +}$, and normalize these operators so that we have the OPE
\be
\sigma_{2,+}(z')\sigma_2^{+}(z)\sim {1\over (z'-z)}
\ee
On the cylinder this implies
\be
\lim_{\tau'\r\infty} \lim_{\tau\r-\infty}{}^{(1)}\langle 0|\otimes {}^{(2)}\langle 0|
e^{\h(\tau'+i\sigma)}\sigma_{2,+}(\tau'+i\sigma)e^{-\h(\tau+i\sigma)}\sigma_2^+(\tau+i\sigma)|0\rangle^{(1)}\otimes |0\rangle^{(2)}=1
\label{qwone}
\ee
where $|0\rangle^{(1)}\otimes |0\rangle^{(2)}
$ is the  NS vacuum of the untwisted sector. 

Let us perform a spectral flow (\ref{spectral}) with parameter $\alpha=-1$. This changes the untwisted NS vacuum as
\be
|0\rangle^{(1)}\otimes |0\rangle^{(2)}\r |0_R^{-}\rangle^{(1)}\otimes |0_R^{-}\rangle^{(2)}
\ee
We also have to ask what happens to the twist operator $\sigma_2^+(w_0)$ under this spectral flow. The action of spectral flow is simple for operators where the fermion content can be expressed as a simple exponential in the language where we bosonise the fermions. For such operators with charge $j$, spectral flow with parameter $\alpha$ leads to a multiplicative factor \cite{acm}
\be
\hat O_j(w)\r e^{-\alpha j w} \hat O_j(w)
\label{spectralformula}
\ee
The operator $\sigma_2^+$ is indeed of this simple form \cite{lm2}, so we just get a multiplicative factor under spectral flow. 
Its charge is $q=\h$, so we get
\be
\sigma_2^+(w)\r e^{-\alpha \h w} \sigma_2^+(w_0)=e^{ \h (\tau+i\sigma)} \sigma_2^+(\tau+i\sigma)
\ee
The operator $\sigma_{2,+}$ has charge $-\h$, so we get
\be
\sigma_{2,+}(w')\r e^{\alpha \h w'} \sigma_{2,+}(w')=e^{ -\h (\tau'+i\sigma)} \sigma_{2,+}(\tau'+i\sigma)
\ee
Thus under spectral flow the relation (\ref{qwone}) gives
\be
\lim_{\tau'\r\infty} \lim_{\tau\r-\infty}{}^{(1)}\langle 0_{R,-}|\otimes {}^{(2)}\langle 0_{R,-}|~
\sigma_{2,+}(\tau'+i\sigma)\sigma_2^+(\tau+i\sigma)~|0_R^-\rangle^{(1)}\otimes |0_R^-\rangle^{(2)}=1
\label{qwonep}
\ee
We write
\be
|0^-_R\rangle\equiv \lim_{\tau\r-\infty}\sigma_2^+(\tau+i\sigma)~|0_R^-\rangle^{(1)}\otimes |0_R^-\rangle^{(2)}
\label{qwtwo}
\ee
This is one of the two  Ramond vacua of the CFT on the double circle. The spin of this vacuum is $-\h$, arising from the spin $-\h$ on each of the two initial Ramond vacua before twisting and the spin $\h$ of the twist operator $\sigma_2^+$. Similarly, we write
\be
\langle 0_{R,-}|\equiv \lim_{\tau'\r\infty} {}^{(1)}\langle 0_{R,-}|\otimes {}^{(2)}\langle 0_{R,-}|~
\sigma_{2,+}(\tau'+i\sigma)
\label{qsone}
\ee
From the relation (\ref{qwonep}) we have
\be
\langle 0_{R,-}|0^-_R\rangle=1
\label{qstwo}
\ee
The relation (\ref{qwtwo}) implies that if we insert $\sigma_2^+$ at a general point $w$, then we will get a state of the form
\be
\sigma_2^+(w)|0_R^-\rangle^{(1)}\otimes |0_R^-\rangle^{(2)}=|0_R^-\rangle+\dots
\label{qwthree}
\ee
where the coefficient of the vacuum on the RHS is unity, and the `$\dots$' represent excited states of the CFT on the doubly wound circle. We will use this relation below.

\section{The amplitude to be computed}

\begin{figure}[ht]
\begin{center}
\includegraphics[width=4cm]{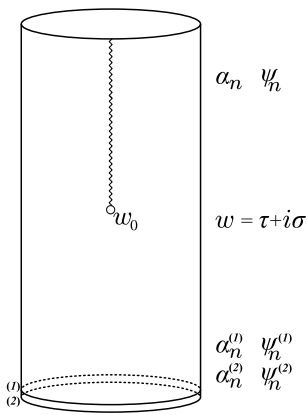}
\caption{Before the twist insertion we have boson and fermion modes on two copies of the $c=6$ CFT. These modes are labeled with superscripts $(1), (2)$ respectively. The twist inserted at $w_0$ joins these to one copy for $\tau>\tau_0$; the modes here do not carry a superscript. The branch cut above $w_0$ indicates that we have two sets of fields at any given $\sigma$; these two sets go smoothly into each other as we go around the cylinder, giving a continuous field on a doubly wound circle.}\label{two}
\end{center}
\end{figure}

Consider the amplitude depicted in figure~.\ref{two}. Let us write down all the states and operators in this amplitude.

\subsection{The initial state}

We have two component strings. Since each is in the Ramond sector, we have to choose one of the Ramond ground states (\ref{rground}). Let us take the state
\be
|\Psi\rangle_i=|0_R^{--}\rangle^{(1)}\otimes |0_R^{--}\rangle^{(2)}
\label{initial}
\ee

\subsection{The final state}

We insert the operator $\hat O_{\dot A\dot B}$ at the point $w_0$ on the cylinder.  Thus the final state that we want to find is given by
\be
|\Psi\rangle_f=\hat O_{\dot A\dot B}(w_0)|\Psi\rangle_i=\Big [{1\over 2\pi i}\int_{w_0} dw G^-_{\dot A}(w)\Big ]\Big [{1\over 2\pi i} \int_{\bar w_0} d\bar w {\bar G}^-_{\dot B}(\bar w)\Big ]~\sigma_2^{++}(w_0)
|0_R^{--}\rangle^{(1)}\otimes |0_R^{--}\rangle^{(2)}
\label{pone}
\ee
The final state will contain one component string with winding number $2$, since the deformation operator contains the twist $\sigma_2$. 

From this stage on, we will write only the left moving part of the state, and join it up with the right moving part at the end. Thus we write
\be
|\Psi\rangle_f=|\psi\rangle|\bar\psi\rangle
\ee
and work with $|\psi\rangle$ in what follows.

\subsection{Outline of the computation}\label{outline}

Let us outline our steps for computing $|\psi\rangle$.

\b

(a) The essence of the computation lies in the nature of the deformation operator. This operator is given by a supercharge acting on the twist operator $\sigma_2^+$.  This supercharge is given by a contour integral of $G^-_{\dot A}$ around the twist insertion. We first deform this contour to a pair of contours: one above and one below the insertion. These contours give zero modes of the supercurrent on the states before and after the twist insertion. We handle these zero modes at the end, and focus first on the state  produced by just the twist insertion $\sigma_2^+$; we call this state $|\chi\rangle$.  

\b

(b) Let us now look at the nature of the twist operator for bosonic fields. As we circle the twist, the two copies of the boson go into each other. The twist operator is defined by cutting a small hole around its insertion $w_0$, and taking boundary condition at the edge of this hole given by filling the hole in the {\it covering space} with a disc; i.e. there are no operator insertions in this covering space and we have just the vacuum state \cite{lm1}. To use this structure of the twist operator, we first map the cylinder to the plane via $z=e^w$, and then pass to the covering space $t$ by the map $z=z_0+t^2$ (here $z_0=e^{w_0}$ is the location of the twist). The small hole cut out  on the cylinder around $w_0$ becomes a small hole around $t=0$. Since the boundary condition on the edge of this hole is generated by filling this hole by a disc, we get just the vacuum state at the origin in the $t$ plane. This observation takes into account the entire effect of the twist on the bosons. 

\b

(c) On the cylinder we can specify the initial state of the system on the two circles at $\tau\r-\infty$ corresponding to the two copies of the $c=6$ CFT. On  the $t$ plane these circles map to punctures at $t=\pm z_0^\h\equiv \pm ia$. Since we have taken no bosonic excitations in our initial state, the bosonic part of the states at these punctures is just the vacuum, and we can close these punctures smoothly, just like the hole at $t=0$. Thus we have no insertions anywhere in the $t$ plane.

\b

(d) Our goal is to find the state at a circle $\tau\r\infty$ on the cylinder. But this circle maps to the circle $|t|=\infty$ on the $t$ plane. Thus what we need is the state in the $t$ plane at infinity.  But since there are no insertions anywhere on the $t$ plane, this state is just the $t$ plane vacuum $|0\rangle_t$. One might think that this means there are no excitations in the final state, but this is not the case: the vacuum on the $t$ plane is killed by positive energy modes defined with respect to the $t$ coordinate, and these will map to a linear combination of positive and negative energy modes in the original cylinder coordinate $w$. Thus all we have to do is express the state $|0\rangle_t$ in terms of the modes on the cylinder, and we would have obtained the bosonic part of the state arising from the twist insertion. 

\b

(e) Let us now ask if we can guess the nature of this state in terms of the modes on the cylinder. In the treatment of quantum fields on curved space we often come across a situation where we have to express the vacuum with respect to one set of field  modes in terms of operators defined with respect to another set of field modes. The state with respect to the latter modes has the form of an exponential of a quadratic, i.e. of the form $e^{\gamma_{mn}a^\dagger_m a^\dagger_n}|0\rangle$. The essential reason for getting this form for the state can be traced to the fact that free fields have a quadratic action, and if there are no operator insertions anywhere then in all descriptions of the state we can only observe exponentials of a quadratic form. 

For our problem, we make the ansatz that the state $|\chi\rangle$ has a similar exponential form. We find the $\gamma_{mn}$ by computing the inner product of the state with a state containing a pair of operator modes. In Appendix \ref{cc} we prove that this exponential ansatz is indeed correct to all orders. Such a form for the state is termed a squeezed state in atomic physics.

\b

(f) Let us now ask if similar arguments can be applied to the fermionic field. The initial state on the cylinder has  Ramond vacua for the two copies of the CFT. If we map to the $t$ plane these would give nontrivial states at $t=\pm ia$. Thus we first perform a spectral flow on the cylinder, which maps the problem to one where these Ramond vacua map to NS vacua at $\tau\r-\infty$ on the cylinder. These NS vacua will map to  NS vacua at $t=\pm ia$, so there will be no operator insertions in the $t$ plane at these points. We should also check the effect of this spectral flow on the twist $\sigma_2^+(w_0)$. 
From (\ref{spectralformula}) we find that  $\sigma_2^+(w_0)$ will change by just a multiplicative constant. This constant will not matter at the end since we know the normalization of our final state by (\ref{qwthree}).

We can now pass to the covering space $t$. We must now ask for the state at the edge of the hole around $t=0$. One finds that the fermions in the $t$ plane are antiperiodic around $t=0$ \cite{lm2}. Thus the state given by the operator $\sigma_2^+$ 
corresponds to having the positive spin Ramond vacuum $|0^+_R\rangle_t$. 
As it stands this implies that we have  a nontrivial state at $t=0$, but we perform another spectral flow, this time in the $t$ plane. Under this second spectral flow the Ramond vacuum $|0^+_R\rangle_t$ maps to the NS vacuum of the $t$ plane $|0\rangle_t$. We take the normalization
\be
{}_t\langle 0|0\rangle _t=1
\label{qsthree}
\ee
for this NS vacuum. At this stage we have indeed no insertions anywhere in the $t$ plane, and the state at $t=\infty$ is just the $t$ plane vacuum for the fermions. Since all coordinate maps and spectral flows were linear in the operator modes, we again expect the state to be given by the exponential of a bilinear in fermion modes. We write such an ansatz, and find the  coefficients in the exponential. 

\b

(g) We can summarize the above discussion in the following general relation
\be
\langle 0_{R,-}|\Big ( \hat O_1, \hat O_2, \dots \hat O_n\Big ) \sigma_2^+(w_0)|0_R^-\rangle^{(1)}\otimes |0_R^-\rangle^{(2)}
={}_t\langle 0|\Big (\hat O'_1, \hat O'_2, \dots \hat O'_n\Big )|0\rangle_t
\label{master}
\ee
The state $\langle 0_{R,-}|$ was defined in (\ref{qsone}). $\hat O_i$ are any operators inserted after the twist insertion (we will need to insert boson and fermion modes in finding the coefficients $\gamma_{mn}$). On the RHS, the operators $\hat O'_i$ are obtained by mapping the operators $\hat O_i$ through all coordinate changes and spectral flows till we reach the $t$ plane with the NS vacuum at $t=0$. The normalizations (\ref{qstwo}) and (\ref{qsthree}) tell us that there is no extra constant relating the LHS to the RHS of (\ref{master}). 

\b

(h) After obtaining the state $|\chi\rangle$ generated by the action of the twist $\sigma_2^+$ on $|0_R^-\rangle^{(1)}\otimes |0_R^-\rangle^{(2)}$, we act with the zero mode of the supercurrent to obtain the final state $|\psi\rangle$ obtained by the action of the full deformation operator on $|0_R^-\rangle^{(1)}\otimes |0_R^-\rangle^{(2)}$.

\section{Coordinate maps and spectral flows}

In this section we carry out the transformations that map the problem to a simple one in the covering $t$ plane. 

\subsection{Mode expansions on the cylinder}

We  expand the fields in modes as follows:

\b

\noindent {\bf Below the twist insertion ($\tau<\tau_0$):}

\b

In this region  we have
\be
\alpha^{(i)}_{A\dot A, n}= {1\over 2\pi} \int_{\sigma=0}^{2\pi} \p_w X^{(i)}_{A\dot A}(w) e^{nw} dw, ~~~i=1,2
\ee
\be
d^{(i)\alpha A}_n={1\over 2\pi i} \int_{\sigma=0}^{2\pi} \psi^{(i)\alpha A}(w) e^{nw} dw, ~~~i=1,2
\ee
which gives
\be
\p_w X^{(i)}_{A\dot A}(w)=-i \sum_n \alpha^{(i)}_{A\dot A, n} e^{-nw} , ~~~i=1,2
\label{pfthree}
\ee
\be
\psi^{(i)\alpha A}(w)=\sum_n d^{(i)\alpha A}_n e^{-n w}, ~~~i=1,2
\ee

The commutation relations are
\be
[\alpha^{(i)}_{A\dot A, m}, \alpha^{(j)}_{B\dot B, n}]=-\epsilon_{AB}\epsilon_{\dot A\dot B}\delta^{ij} m \delta_{m+n,0}
\ee
\be
\{ d^{(i)\alpha A}_m, d^{(j)\beta B}_n\}=-\epsilon^{\alpha\beta}\epsilon^{AB}\delta^{ij}\delta_{m+n,0}
\ee

\b

\noindent {\bf Above the twist insertion ($\tau>\tau_0$):}

\b

In this region we have a doubly twisted circle. We have a choice of normalization in how we define modes on the doubly wound circle, and we take
\be
\alpha_{A\dot A, n}= {1\over 2\pi} \int_{\sigma=0}^{4\pi} \p_w X_{A\dot A}(w) e^{{n\over 2}w} dw
\label{qaone}
\ee
\be
d^{\alpha A}_n= {1\over 2\pi i}\int_{\sigma=0}^{4\pi} \psi^{\alpha A}(w) e^{{n\over 2} w} dw
\ee
 Taking the normalizations as above, one finds
\be
\p_w X_{A\dot A}(w)=-\h i \sum_n \alpha_{A\dot A, n} e^{-{n\over 2}w}
\ee
\be
\psi^{\alpha A}(w)=\h \sum_n d^{\alpha A}_n e^{-{n\over 2} w}
\ee

Note the  factor of $\h$ that appears in these  equations.  
The commutation relations turn out to be
\be\label{bcommtwist}
[\alpha_{A\dot A, m}, \alpha_{B\dot B, n}]=-\epsilon_{AB}\epsilon_{\dot A\dot B} m \delta_{m+n,0}
\ee
\be\label{fcommtwist}
\{ d^{\alpha A}_m, d^{\beta B}_n\}=-2\epsilon^{\alpha\beta}\epsilon^{AB}\delta_{m+n,0}
\ee
Again note the factor of $2$ in the fermion relation. The difference between the boson and the fermion cases arises from the fact that they have different scaling dimensions.

\subsection{The $G^-_{\dot A, -\h}$ operator}

\begin{figure}[ht]
\begin{center}
\includegraphics[width=8cm]{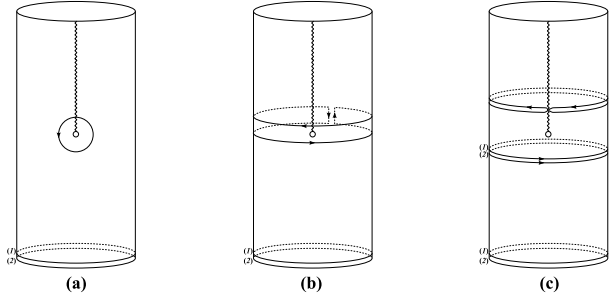}
\caption{(a) The supercharge in the deformation operator is given by integrating $G^-_{\dot A}$ around the insertion at $w_0$. (b) We can stretch this contour as shown, so that we get a part above the insertion and a part below, joined by vertical segments where the contributions cancel. (c) The part above the insertion gives the zero mode of the supercharge on the doubly wound circle, while the parts below give the sum of this zero mode  for each of the two initial copies of the CFT.}\label{G-contour}
\end{center}
\end{figure}

Let us first put the ${1\over 2\pi i}\int_{w_0} dw G^-_{\dot A}(w)$ operator in (\ref{pone}) in a more convenient form. The contour in this operator runs circles the insertion $w_0$ (figure~\ref{G-contour}(a)). We  can stretch this to a contour that runs around the rectangle shown in figure~\ref{G-contour}(b). The vertical sides of the contour cancel out.  
We can thus break the contour into a part above the insertion and a part below the insertion (fig.~\ref{G-contour}(c)). The lower leg gives 
\be
{1\over 2\pi i} \int_{w=\tau_0-\epsilon}^{\tau_0-\epsilon+2\pi i}G^-_{\dot A}(w)={1\over 2\pi i} \int_{w=\tau_0-\epsilon}^{\tau_0-\epsilon+2\pi i}[G^{(1)-}_{\dot A}(w)+G^{(2)-}_{\dot A}(w)]dw\equiv G^{(1)-}_{\dot A, 0}+G^{(2)-}_{\dot A, 0}
\ee
 The upper leg gives
\be
-{1\over 2\pi i} \int_{w=\tau_0+\epsilon}^{\tau_0+\epsilon+4\pi i}G^-_{\dot A}(w)dw\equiv -G^-_{\dot A, 0}
\ee
where we note that the two copies of the CFT have linked into one  copy on a doubly wound circle, and we just get the zero mode of $G^-_{\dot A}$ on this single copy.

Note that
\be
G^{(i)-}_{\dot A, 0}|0_R^{--}\rangle^{(i)}=0, ~~~i=1,2
\ee
since the $\alpha$ index of $|0_R^{\alpha -}\rangle^{(i)}$ forms a doublet under $SU(2)_L$, and 
we cannot further lower the spin of $|0_R^{--}\rangle^{(i)}$ without increasing the energy level. Thus the lower contour gives nothing, and we have
\be
|\psi\rangle=-G^-_{\dot A, 0}\sigma_2^{+}(w_0)
|0_R^{-}\rangle^{(1)}\otimes |0_R^{-}\rangle^{(2)}
\label{ptwo}
\ee

Let us write this as 
\be
|\psi\rangle=-G^-_{\dot A, 0}|\chi\rangle
\ee
where
\be
|\chi\rangle=\sigma_2^{+}(w_0)
|0_R^{-}\rangle^{(1)}\otimes |0_R^{-}\rangle^{(2)}
\ee

\subsection{Ansatz for $|\chi\rangle$}

Let us leave aside the action of $G^-_{\dot A, 0}$ for the moment, and consider the state $|\chi\rangle$.

The spin of  $|\chi\rangle$ is $-\h$, since each of the two Ramond ground states have spin $-\h$ and $\sigma_2^+$ has spin $\h$. Further, the fermions on the double circle produced after the twist are periodic after we go around the double circle; this follows by noting that these fermions were   periodic on each of the two copies before the twist, and we have simply cut and rejoined the copies into one long loop. Thus the state $|\chi\rangle$ can be considered as built up by adding excitations (with total charge zero) to the Ramond vacuum $|0^-_R\rangle$ of the doubly twisted theory (we assume that this vacuum is normalized to unit norm)
\be
|\chi\rangle=\prod \{ \alpha_{C\dot C, m_i}\} \prod \{ \psi^{\beta}_{\dot D, n_j}\} |0^-_R\rangle
\label{pfour}
\ee
As discussed in section (\ref{outline}), the state $|\chi\rangle$ should have  the form of an exponential in the boson and fermion creation operators. Let us write down the ansatz and then explain some of its points. In Appendix \ref{cc} we show that for the bosonic case this ansatz is correct to all orders in the number of excitations; the fermionic case can be done in a similar way. We will write
\be
|\chi\rangle=e^{\sum_{ m\ge 1, n\ge 1}\gamma^B_{mn}[-\alpha_{++, -m}\alpha_{--, -n}+\alpha_{-+, -m}\alpha_{+-, -n}]}
e^{\sum_{m\ge0,n\ge 1}\gamma^F_{mn}[d^{++}_{-m}d^{--}_{-n}-d^{+-}_{-m}d^{-+}_{-n}]}|0^-_R\rangle
\label{pfive}
\ee
Below we will define more precisely the modes $\alpha_{A\dot A, n}, d^{\alpha A}_n$ on the double circle. For now, Let us note some points about the above expression:

\b

(a) From eq. (\ref{qwthree}) we see that there will not be any additional multiplicative constant on the RHS; the coefficient of first term obtained by expanding the exponential (i.e. the number unity) is set by (\ref{qwthree}).

\b

(b) The initial state $|0^-_R\rangle$ is a singlet of $SO(4)_I$, the symmetry group (\ref{pthree}) in the torus directions. The operator $\sigma^+_2$ is a singlet under this group also. Thus the state $|\chi\rangle$ will have to be a singlet under this group. Thus we have written the ansatz in a way that the $A, \dot A$ indices of $\alpha_{A\dot A}, \psi^{\alpha A}$ are grouped to make singlets under $SO(4)_I$.

\b

(c) We have 
\be
\alpha_{A\dot A, 0}|0^-_R\rangle=0
\ee
since there is no momentum in the state. Thus the mode summations for the bosons start with $m,n=1$ and not with $m,n=0$. 

\b

(d) We have
\be
d_0^{-A}|0^-_R\rangle=0
\ee
Thus the sum over fermion modes starts with $n=1$ for $d^{--}, d^{-+}$ and with $m=0$ for $d^{++}, d^{+-}$.  By writing modes this way we remove a normal ordering term that can arise from the anticommutation of zero modes; such a contribution would then have to be reabsorbed in an overall normalization constant in front of the exponential. We will find later that the value $m=0$ does not occur either because the $\gamma^F_{mn}$ vanish for that case; in fact we will find that the $\gamma^B_{mn}, \gamma^F_{mn}$ are nonzero only for $m,n$ odd.

\subsection{The first spectral flow}

Let us continue to work with the state $|\chi\rangle$, and return to $|\psi\rangle$ at the end. 

First we perform a spectral flow (\ref{spectral}) with parameter $\alpha=1$. Let the resulting state be called $|\chi\rangle_{\alpha=1}$. This spectral flow has the following effects:

\b

(a) The Ramond ground states $|0_R^{-}\rangle^{(i)}$, $i=1,2$ change to NS vacua
\be
|0_R^{-}\rangle^{(1)}\r |0\rangle^{(1)}, ~~~|0_R^{-}\rangle^{(2)}\r |0\rangle^{(2)}
\ee

\b

(b) The operator $\sigma_2^+$ changes by a phase which depends on its charge $q$; this charge is $q=\h$. So we get
\be
\sigma_2^+(w_0)\r e^{-\h w_0} \sigma_2^+(w_0)
\ee

Thus we get
\be
|\chi\rangle_{\alpha=1}=e^{-\h w_0}\sigma_2^+(w_0)|0\rangle^{(1)}\otimes |0\rangle^{(2)}
\ee

\b

(c) Modes of bosonic operators are not affected. The fermionic field changes as follows
\be
\psi^{+ A}(w)\r e^{-\h w} \psi^{+A}(w), ~~~ \psi^{- A}(w)\r e^{\h w} \psi^{-A}(w)
\label{pseven}
\ee
Thus fermionic modes change as follows
\be
d^{(i)\pm A}_n\r {1\over 2\pi i} \int_{\sigma=0}^{2\pi} \psi^{(i)\pm A}(w) e^{(n\mp\h)w} dw, ~~~i=1,2
\ee
\be
d^{\pm A}_n\r  {1\over 2\pi i}\int_{\sigma=0}^{4\pi} \psi^{\pm A}(w) e^{{(n\mp 1)\over 2} w} dw
\label{psix}
\ee

\b

(d) For $\tau>\tau_0$, we have one copy of the CFT on a doubly long circle. Before the spectral flow, the state in this region was built by applying excitations to $|0^-_R\rangle$ (eq. (\ref{pfour})). Under the spectral flow we have
\be
|0^-_R\rangle\r |0^+_R\rangle
\ee

\subsection{Mode expansions on the $z$ plane}

We wish to go to a covering space which will allow us to see explicitly the action of the twist operator. As a preparatory step, it is convenient to map the cylinder with coordinate $w$ to the plane with coordinate $z$
\be
z=e^w
\ee
Under this map the operator modes change as follows. Before the insertion of the twist ($|z|<e^{\tau_0}$) we have
\be
\alpha^{(i)}_{A\dot A, n}\r {1\over 2\pi} \int_{z=0} \p_z X^{(i)}_{A\dot A}(z) z^n dz, ~~~i=1,2
\ee
\be
d^{(i)+ A}_n={1\over 2\pi i} \int_{z=0} \psi^{(i)+ A}(z) z^{n-1} dz, ~~~i=1,2
\ee
\be
d^{(i)- A}_n\r{1\over 2\pi i} \int_{z=0} \psi^{(i)- A}(z) z^{n} dz, ~~~i=1,2
\ee
After the twist ($|z|>e^{\tau_0}$) we have
\be
\alpha_{A\dot A, n}\r {1\over 2\pi} \int_{z=\infty} \p_z X_{A\dot A}(z) z^{{n\over 2}} dz
\ee
\be
d^{+ A}_n\r {1\over 2\pi i}\int_{z=\infty} \psi^{+ A}(z) z^{{(n-2)\over 2} } dz
\ee
\be
d^{- A}_n\r {1\over 2\pi i}\int_{z=\infty} \psi^{- A}(z) z^{{n\over 2} } dz
\ee

\subsection{Mapping to the covering space}

\begin{figure}[ht]
\begin{center}
\includegraphics[width=5cm]{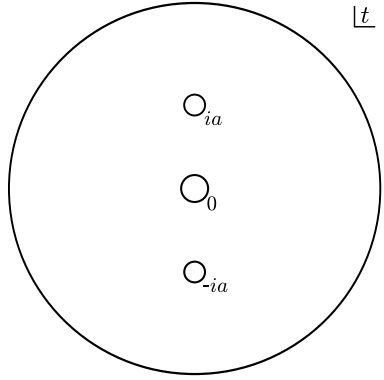}
\caption{The $z$ plane is mapped to the covering space -- the $t$ plane -- by the map $z=z_0+t^2$. The point $z=0$ corresponds to $\tau\r-\infty$ on the cylinder, and the two copies of the CFT there correspond to the points $t=\pm ia$. The location of the twist operator maps to $t=0$. The top the cylinder $\tau\r\infty$ maps to $t\r \infty$. After all maps and spectral flows, we have the NS vacuum at $t=0, \pm ia$, and so we can smoothly close all these punctures. The state $|\chi\rangle$ is thus just the $t$ plane vacuum; we must write this in terms of the original cylinder modes and apply the supercharge to get the final state $|\psi\rangle$.}\label{t-plane}
\end{center}
\end{figure}

We pass to the cover of the $z$ plane via the map
\be
z=z_0+t^2
\ee
where
\be
z_0=e^{w_0}\equiv a^2
\ee
Thus the map from the $z$ plane to the $t$ plane has second order branch points at $z=z_0$ (the location of the twist $\sigma_2^+$) and at infinity (corresponding to the top of the cylinder, where we can imagine the dual twist $\sigma_{2,+}$ being placed). Under this map we have the following changes

\b

(a) Consider the circle on the cylinder at $\tau\r -\infty$; this is  the location of the initial states on the cylinder. This circle maps to $z=0$, and then to $t=\pm ia$ on the $t$ plane. There were two copies of the CFT at $\tau\r-\infty$ on the cylinder, and the initial state of one copy (copy $(1)$) will map to the point $t=ia$ and the state of the other copy (copy $(2)$) will map to $t=-ia$. 

Note that in our present problem the states of  these copies, 
were $|0^-_R\rangle^{(i)}$ to start with, which became NS vacua $|0\rangle^{(i)}$ after the first spectral flow. Now when we map them to the $t$ plane we find that there is just the $t$ plane vacuum at the points $\pm ia$, so we may smoothly close the punctures at these points with no insertions at the puncture.  

\b

(b) The location of the twist insertion $\sigma_2^+$ maps to $t=0$. At this location we have the state $|0^+_R\rangle_t$, the spin up Ramond vacuum of the $t$ plane.

\b

(c) The operator modes become before the twist
\be
\alpha^{(1)}_{A\dot A, n}\r {1\over 2\pi} \int_{t=ia} \p_t X_{A\dot A}(t) (z_0+t^2)^n dt
\ee
\be
\alpha^{(2)}_{A\dot A, n}\r {1\over 2\pi} \int_{t=-ia} \p_t X_{A\dot A}(t) (z_0+t^2)^n dt
\ee
\be
d^{(1)+ A}_n\r {1\over 2\pi i} \int_{t=ia} \psi^{+ A}(t) (z_0+t^2)^{n-1}(2t)^\h dt
\ee
\be
d^{(2)+ A}_n\r {1\over 2\pi i} \int_{t=-ia} \psi^{+ A}(t) (z_0+t^2)^{n-1}(2t)^\h dt
\ee
\be
d^{(1)- A}_n\r {1\over 2\pi i} \int_{t=ia} \psi^{- A}(t) (z_0+t^2)^{n} (2t)^\h dt
\ee
\be
d^{(2)- A}_n\r{1\over 2\pi i} \int_{t=-ia} \psi^{- A}(t) (z_0+t^2)^{n} (2t)^\h dt
\ee
After the twist we have
\be
\alpha_{A\dot A, n}\r {1\over 2\pi} \int_{t=\infty} \p_t X_{A\dot A}(t) (z_0+t^2)^{{n\over 2}} dt
\ee
\be
d^{+ A}_n\r {1\over 2\pi i}\int_{t=\infty} \psi^{+ A}(t) (z_0+t^2)^{{(n-2)\over 2} } (2t)^\h dt
\ee
\be
d^{- A}_n\r {1\over 2\pi i}\int_{t=\infty} \psi^{- A}(t) (z_0+t^2)^{{n\over 2} } (2t)^\h dt
\ee

\subsection{The second spectral flow}

We have now mapped the problem to the $t$ plane, where we have found a state $|0^+_R\rangle_t$ at $t=0$, and no other insertions anywhere. Let us perform a spectral flow with $\alpha=-1$ in the $t$ plane. This will have the following effects

\b

(a) The Ramond ground state at $t=0$ maps to the NS vacuum in the $t$ plane
\be
|0^+_R\rangle_t\r |0\rangle_t
\ee

\b

(b) The operator modes change as follows. The bosons are not affected. The fermion field changes as
\be
\psi^{\pm A}(t) \r t^{\pm \h} \psi^{\pm A}(t)
\ee
Before the twist we get the modes
\be
\alpha^{(1)}_{A\dot A, n}\r {1\over 2\pi} \int_{t=ia} \p_t X_{A\dot A}(t) (z_0+t^2)^n dt
\ee
\be
\alpha^{(2)}_{A\dot A, n}\r {1\over 2\pi} \int_{t=-ia} \p_t X_{A\dot A}(t) (z_0+t^2)^n dt
\ee
\be
d^{(1)+ A}_n\r {2^\h\over 2\pi i} \int_{t=ia} \psi^{+ A}(t) (z_0+t^2)^{n-1}t dt
\ee
\be
d^{(2)+ A}_n\r {2^\h\over 2\pi i} \int_{t=-ia} \psi^{+ A}(t) (z_0+t^2)^{n-1}t dt
\ee
\be
d^{(1)- A}_n\r{2^\h\over 2\pi i} \int_{t=ia} \psi^{- A}(t) (z_0+t^2)^{n}  dt
\ee
\be
d^{(2)- A}_n\r{2^\h\over 2\pi i} \int_{t=-ia} \psi^{- A}(t) (z_0+t^2)^{n}  dt
\ee
After the twist we have
\be
\alpha_{A\dot A, n}\r {1\over 2\pi} \int_{t=\infty} \p_t X_{A\dot A}(t) (z_0+t^2)^{{n\over 2}} dt
\label{qatwo}
\ee
\be
d^{+ A}_n\r {2^\h\over 2\pi i}\int_{t=\infty} \psi^{+ A}(t) (z_0+t^2)^{{(n-2)\over 2} } t dt
\ee
\be
d^{- A}_n\r {2^\h\over 2\pi i}\int_{t=\infty} \psi^{- A}(t) (z_0+t^2)^{{n\over 2} }  dt
\ee

In the $t$ plane, we will also define mode operators that are natural to the $t$ plane. Thus we write
\be
\t\alpha_{A\dot A, n}\equiv  {1\over 2\pi} \int_{t=0} \p_t X_{A\dot A}(t) t^n dt
\label{qathree}
\ee
\be
\t d^{\alpha A}_r= {1\over 2\pi i}\int_{t=0} \psi^{\alpha A}(t) t^{r-\h} dt
\ee
Note that the bosonic index $n$ is an integer while the fermionic index $r$ is a half integer.  We have
\be
\t\alpha_{A \dot A,m}|0\rangle_t=0, ~~~m\ge 0
\label{ptthree}
\ee
\be
\t d^{\alpha A}_r|0\rangle_t=0, ~~~r>0
\label{ptthreep}
\ee
The commutation relations are
\be
[\t\alpha_{A \dot A}, \t\alpha_{B \dot B}]=-\epsilon_{A B} \epsilon_{\dot A \dot B} m\delta_{m+n, 0}
\label{pttwo}
\ee
\be
\{\t d^{\alpha A}_r, \t d^{\beta B}_s\}=-\epsilon^{\alpha\beta}\epsilon^{AB}\delta_{r+s, 0}
\label{pttwop}
\ee

\section{Computing $\gamma^B_{mn}, \gamma^F_{mn}$}

In this section we compute the  coefficients $\gamma^B_{mn}, \gamma^F_{mn}$. For this computation we use the relation (\ref{master}) to relate correlators of  operators on the cylinder to correlators on the $t$ plane. The latter correlators can be computed very simply, and we thereby find the  coefficients $\gamma^B_{mn}, \gamma^F_{mn}$.

\subsection{The bosonic coefficients $\gamma^B_{mn}$}

Let us compute
\be
\langle 0_{R,-}|\Big (\alpha_{++,_n}\alpha_{--,_m}\Big )\sigma^+_2(w_0)|0_R^-\rangle^{(1)}\otimes |0_R^-\rangle^{(2)}={}_t\langle 0 |\Big (\alpha'_{++,_n}\alpha'_{--,_m}\Big )|0\rangle_t
\label{ptone}
\ee
where the primes on the operators on the RHS signify the fact that these operators arise from the unprimed operators by the various maps leading to the spectral flowed $t$ plane description. 
Using the ansatz (\ref{pfive}) we can write the LHS as
\bea
&&\langle 0_{R,-}|\Big (\alpha_{++,_n}\alpha_{--,_m}\Big ) \nn
&& \qquad \times  e^{\sum_{ m\ge 1, n\ge 1}\gamma^B_{mn}(-\alpha_{++, -m}\alpha_{--, -n}+\alpha_{-+, -m}\alpha_{+-, -n})}
e^{\sum_{m\ge0,n\ge 1}\gamma^F_{mn}(d^{++}_{-m}d^{--}_{-n}-d^{+-}_{-m}d^{-+}_{-n})}|0^-_R\rangle \nn
\eea
Expanding the exponential, and using the commutation relations (\ref{bcommtwist}), one finds that this LHS equals
\be
-mn\gamma^B_{mn}~\langle 0_{R,-}|0^-_R\rangle=-mn\gamma^B_{mn}
\label{ptsix}
\ee
where we have used (\ref{qstwo}). 

The RHS of (\ref{ptone}) can be written by expressing the inserted operators as contour integrals over circles at large $t$
\be
{}_t\langle 0 |\Big ({1\over 2\pi i}\int_{} dt_1 \p_t X_{++}(t_1) (z_0+t_1^2)^{{n\over 2}}\Big )\Big(
{1\over 2\pi i}\int_{} dt_2 \p_t X_{--}(t_2) (z_0+t_2^2)^{{m\over 2}}\Big )
|0\rangle_t
\label{ptfive}
\ee
with $|t_1|>|t_2|$. We have\footnote{The symbol ${}^nC_m$ is the binomial coefficient, also written as $\pmatrix {n\cr m\cr }$.}
\be
(z_0+t_1^2)^{n\over 2}=t_1^n[1+z_0t_1^{-2}]^{n\over 2}=\sum_{p\ge 0}{}^{n/2}C_p ~z_0^p ~t_1^{n-2p}
\ee
\be
(z_0+t_2^2)^{m\over 2}=t_2^m[1+z_0t_2^{-2}]^{m\over 2}=\sum_{q\ge 0}{}^{m/2}C_q ~z_0^q ~t_2^{m-2q}
\ee
Equating (\ref{ptsix}) and (\ref{ptfive}) gives
\be
\gamma_{mn}=-{1\over mn}\sum_{p\ge 0} \sum_{q\ge 0}~{}^{n/2}C_p~{}^{m/2}C_q~ z_0^{p+q} {}_t\langle 0|\t \alpha_{++,n-2p} \t \alpha_{--,m-2q}|0\rangle_t
\ee
Using the commutation relations (\ref{pttwo}) we get
\be
m-2q=-(n-2p) ~\Rightarrow ~q={m+n\over 2}-p
\ee
Since $p,q$ are  integral,  either $n,m$ are both even or both odd.
Using Eq. (\ref{ptthree}) we find
\be
\gamma_{mn}={1\over mn}\sum_{p= 0} ^{[{n\over 2}]}~{}^{n/2}C_p~{}^{m/2}~C_{{m+n\over 2}-p} ~z_0^{m+n\over 2} (n-2p)
\ee
where the symbol $[~]$ stands for `floor' (i.e. `integer part of'). 
We can perform this sum using a symbolic manipulation program like {\it Mathematica}\footnote{The sum can be evaluated by hand by writing it in terms of the series representation of ${}_2F_1(-\frac{m-1}{2},1;\frac{n-1}{2}+2;1)$.}. 
 With $n,m$ both even we get
\be
\gamma^B_{mn}=0
\ee
For $n,m$ odd we write
\be
m=2m'+1, ~~~n=2n'+1
\ee 
and find
\be\label{gammaB}
\gamma^B_{2m'+1, 2n'+1}={2\over (2m'+1)(2n'+1)} {z_0^{(1+m'+n')} \Gamma[{3\over 2}+m']\Gamma[{3\over 2}+n']\over (1+m'+n')\pi \Gamma[m'+1]\Gamma[n'+1]}
\ee

\vskip .5 in 

\subsection{The fermionic coefficients $\gamma^F_{mn}$}

We follow the same steps to find $\gamma^F_{mn}$.
Let us compute
\be
\langle 0_{R,-}|\Big (d^{++}_nd^{--}_m\Big )\sigma^+_2(w_0)|0_R^-\rangle^{(1)}\otimes |0_R^-\rangle^{(2)}={}_t\langle 0 |\Big (d'^{++}_nd'^{--}_m\Big )|0\rangle_t
\label{ptonep}
\ee
We can write the LHS as
\bea
&&\langle 0_{R,-}|[d^{++}_nd^{--}_m] \nn
&& \qquad \times e^{\sum_{ m\ge 1, n\ge 1}\gamma^B_{mn}[-\alpha_{++, -m}\alpha_{--, -n}+\alpha_{-+, -m}\alpha_{+-, -n}]}
e^{\sum_{m\ge0,n\ge 1}\gamma^F_{mn}[d^{++}_{-m}d^{--}_{-n}-d^{+-}_{-m}d^{-+}_{-n}]}|0^-_R\rangle \nn
\eea
Expanding the exponential, and using the commutation relations (\ref{fcommtwist}), one finds that this LHS equals
\be
4\gamma^F_{mn}\langle 0_{R,-}|0^-_R\rangle=4\gamma^F_{mn}
\label{ptsixp}
\ee
The RHS of  (\ref{ptonep}) gives
\be
{}_t\langle 0 |\Big [{2^\h\over 2\pi i}\int_{} dt_1\psi^{++}(t_1) (z_0+t_1^2)^{{n-2\over 2}}t_1\Big ]\Big[
{2^\h\over 2\pi i}\int_{} dt_2\psi^{--}(t_2) (z_0+t_2^2)^{{m\over 2}}\Big ]
|0\rangle_t
\label{ptfivep}
\ee
with $|t_1|>|t_2|$. We have
\be
(z_0+t_1^2)^{(n-2)\over 2}=t_1^{(n-2)}[1+z_0t_1^{-2}]^{n-2\over 2}=\sum_{p\ge 0}{}^{(n-2)/2}C_p~ z_0^p ~t_1^{n-2-2p}
\ee
\be
(z_0+t_2^2)^{m\over 2}=t_2^{m}[1+z_0t_2^{-2}]^{m\over 2}=\sum_{q\ge 0}{}^{m/2}C_q ~z_0^q ~t_2^{m-2q}
\ee
Equating (\ref{ptsixp}) and (\ref{ptfivep}) gives
\be
\gamma^{F}_{mn}=\h\sum_{p\ge 0}\sum_{q\ge 0}~{}^{(n-2)/2}C_p ~{}^{m/2}C_q~  {}_t\langle 0| \t d^{++}_{n-2p-{1\over 2}} \t d^{--}_{m-2q+{1\over 2}}|0\rangle_t
\ee
Using the commutation relations (\ref{pttwop}) we get
\be
m-2q+\h=-(n-2p-\h)~ \Rightarrow ~q={m+n\over 2}-p
\ee
From (\ref{ptthreep}) we have that
\be
n-2p-\h>0, ~~~m-2q+\h<0
\ee
Thus we get
\be
\gamma^{F}_{mn}=-\h\sum_{p= 0}^{[{n-1\over 2}]}{}^{(n-2)/2}C_p ~{}^{m/2}C_{{m+n\over 2}-p}~ z_0^{m+n\over 2}
\ee
 With $n,m$ both even we get
\be
\gamma^F_{mn}=0
\ee
For $n,m$ odd we write
\be
m=2m'+1, ~~~n=2n'+1
\ee 
and find
\be\label{gammaF}
\gamma^{F}_{2m'+1, 2n'+1}=-{ z_0^{(1+m'+n')}\Gamma[{3\over 2}+m']\Gamma[{3\over 2}+n']\over (2n'+1)\pi(1+m'+n') \gamma[m'+1]\Gamma[n'+1]}
\ee

\vskip .5 in

\section{The state $|\psi\rangle$}

Finally we return to the computation of $|\psi\rangle$
\be
|\psi\rangle=-G^-_{\dot A, 0}|\chi\rangle
\label{psip}
\ee

\subsection{Applying the supercharge}

We have
\be
G^-_{\dot A, 0}={1\over 2\pi i} \int_{w=\tau}^{w=\tau+4\pi i} G^-_{\dot A}(w) dw = -{i\over 2} \sum_{n=-\infty}^\infty d^{-A}_n \alpha_{A\dot A, -n}
\ee
The positive index modes in the above expression can act on the exponential in $|\chi\rangle$, generating negative index modes. We would like to write $|\psi\rangle$ with only negative index modes acting on $|0^-_R\rangle$; these modes have trivial commutation and anticommutation relations with each other. Thus write
\be
G^-_{\dot A, 0}=-{i\over 2} \sum_{n>0}^\infty d^{-A}_{-n} \alpha_{A\dot A, n}
-{i\over 2} \sum_{n>0}^\infty d^{-A}_{n} \alpha_{A\dot A, -n}
\ee
Recall that $\gamma^B_{mn}, \gamma^F_{mn}$ are nonzero only for odd indices. Thus we write $n=2n'+1, m=2m'+1$, and find
\be
-{i\over 2} \sum_{n>0}^\infty d^{-A}_{-n} \alpha_{A\dot A, n}|\chi\rangle=-{i\over 2}\sum_{n'\ge 0, m'\ge 0} (2n'+1) \gamma^B_{2m'+1, 2n'+1} d^{-A}_{-(2n'+1)} \alpha _{A\dot A, -(2m'+1)}|\chi\rangle
\ee
\be
-{i\over 2} \sum_{n>0}^\infty d^{-A}_{n} \alpha_{A\dot A, -n}|\chi\rangle=i \sum_{n'\ge 0, m'\ge 0}\gamma^F_{2m'+1, 2n'+1} d^{-A}_{-(2n'+1)}\alpha_{A\dot A, -(2m'+1)}|\chi\rangle
\ee
Thus
\be
G^-_{\dot A, 0}|\chi\rangle=-i\sum_{n'\ge 0, m'\ge 0}\Big ((n'+\h)\gamma^B_{2m'+1, 2n'+1}-\gamma^F_{2m'+1, 2n'+1}\Big )d^{-A}_{-(2n'+1)} \alpha_{A\dot A, -(2m'+1)}|\chi\rangle
\ee
Using the values of $\gamma^B, \gamma^F$ from (\ref{gammaB}),(\ref{gammaF}), we find
\bea
(n'+\h)\gamma^B_{2m'+1, 2n'+1}-\gamma^F_{2m'+1, 2n'+1}&&\nonumber \\
=z_0^{m'+n'+1}&&{\Gamma[{3\over 2}+m']\Gamma[{3\over 2}+n']\over (m'+n'+1)\pi \Gamma[m'+1]\Gamma[n'+1]}\Big ( {1\over 2m'+1}+{1\over 2n'+1}\Big )\nonumber\\
=z_0^{m'+n'+1}&&{2\Gamma[{3\over 2}+m']\Gamma[{3\over 2}+n']\over \pi (2m'+1)(2n'+1)\Gamma[m'+1]\Gamma[n'+1]}
\eea
We observe that  the $n'$ and $m'$ sums factorize. We get
\be
G^-_{\dot A, 0}|\chi\rangle=-i\Big (\sum_{n'\ge 0} {2^\h z_0^{n'+\h}\Gamma[{3\over 2}+n'] \over \pi^\h (2n'+1)\Gamma[n'+1]}d^{-A}_{-(2n'+1)}\Big )
\Big (\sum_{m'\ge 0} {2^\h z_0^{m'+\h}\Gamma[{3\over 2}+m'] \over \pi^\h (2m'+1)\Gamma[m'+1]}\alpha_{A\dot A, -(2m'+1)}\Big )|\chi\rangle
\ee

\subsection{The final state}

Finally, we write down the complete final state
\be
|\Psi\rangle_f=|\psi\rangle\otimes |\bar\psi\rangle
\ee
\bea
|\psi\rangle&=&-G^-_{\dot A, 0}|\chi\rangle\nonumber\\
&=&i\Big (\sum_{n'\ge 0} {2^\h z_0^{n'+\h}\Gamma[{3\over 2}+n'] \over \pi^\h (2n'+1)\Gamma[n'+1]}d^{-A}_{-(2n'+1)}\Big )
\Big (\sum_{m'\ge 0} {2^\h z_0^{m'+\h}\Gamma[{3\over 2}+m'] \over \pi^\h (2m'+1)\Gamma[m'+1]}\alpha_{A\dot A, -(2m'+1)}\Big )\nonumber \\
&&e^{\sum_{ p'\ge 0, q'\ge 0}\gamma^B_{2p'+1, 2q'+1}[-\alpha_{++, -(2p'+1)}\alpha_{--, -(2q'+1)}+\alpha_{-+, -(2p'+1)}\alpha_{+-, -(2q'+1)}]}\nonumber \\
&&e^{\sum_{p''\ge 0,q''\ge 0}\gamma^F_{2p''+1, 2q''+1}[d^{++}_{-(2p''+1)}d^{--}_{-(2q''+1)}-d^{+-}_{-(2p''+1)}d^{-+}_{-(2q''+1)}]}~~|0^-_R\rangle
\label{finalstate}
\eea
with a similar expression for $|\bar\psi\rangle=-\bar G^-_{\dot B, 0}|\bar\chi\rangle$.

\section{Discussion}

The D1D5 system has been very useful in the study of black holes. The
entropy of the extremal holes  agrees
(at leading order) with the entropy at the orbifold point \cite{counting}. The agreement extends to near extremal holes, and the
emission rate from the excited CFT state at the orbifold point agrees
with the emission from the near extremal black
hole \cite{radiation}. Energy levels of simple CFT
states at the orbifold point agree with the excitation spectrum of the
corresponding microstate geometries, and radiation from specific black hole microstates agrees exactly with the expected radiation from the corresponding CFT state~\cite{fuzzballs,acm,microrad}.

But to proceed further with the study of microstates we need to study the CFT away from the orbifold point, since the gravity solutions relevant to black hole physics do not correspond to the orbifold point. We can move away from the orbifold point by adding the deformation operator to the Lagrangian of the theory. In this paper we have computed the effect of this deformation operator on the simplest state of the CFT: two `singly wound' copies of the $c=6$ CFT, each in the Ramond ground state $|0^-_R\rangle$. We argued that the effect of the twist operator  should be given by an expression that is of the form of a squeezed state, and we found the  coefficients in the exponential that describe this state. We also have a supercharge applied to this twist operator, and after taking this into account we found the final state given in (\ref{finalstate}).

To apply this expression to a physical problem we have to integrate the location of the deformation operator over the $\tau, \sigma$ space of the 1+1 dimensional CFT, and also consider more general initial states. We will discuss these steps elsewhere, but from the expression (\ref{finalstate}) we can already see some aspects of the general physics: the deformation operator can generate an arbitrary number fermion and boson pairs (besides the fermion and boson supplied by the supercharge).  The falloff with energy of these excitations are given by the expressions for $\gamma^B_{mn}, \gamma^F_{mn}$; this falloff is a power law rather than an exponential. Thus we can have a long `tail' in the distribution with an ever larger number of particle pairs. 

\section*{Acknowledgements}

We thank Justin David for several helpful discussions. We also thank Sumit Das,  Antal Jevicki, Yuri Kovchegov, Oleg Lunin and  Emil Martinec for many helpful comments. The work of SGA and SDM is
supported in part by DOE grant DE-FG02-91ER-40690. The work of BDC was supported by the Foundation for Fundamental Research on Matter.

\appendix
\renewcommand\theequation{\thesection.\arabic{equation}}
\setcounter{equation}{0}

\section{Notation and the CFT algebra} \label{ap:CFT-notation}

We have 4 real left moving fermions $\psi^1, \psi^2, \psi^3, \psi^4$ which we group into doublets $\psi^{\alpha A}$
\be
\pmatrix{\psi^{++}\cr \psi^{-+}\cr}=\sqi\pmatrix{\psi_1+i\psi_2\cr \psi_3+i\psi_4\cr}
\ee
\be
\pmatrix{\psi^{+-}\cr \psi^{--}\cr}=\sqi\pmatrix{\psi_3-i\psi_4\cr -(\psi_1-i\psi_2)\cr}
\ee
Here $\alpha=(+,-)$ is an index of the subgroup $SU(2)_L$ of rotations on $S^3$ and $A=(+,-)$ is an index of the subgroup $SU(2)_1$ from rotations in $T^4$. We have the reality conditions
\be
 (\psi^\dagger)_{\alpha A}=-\epsilon_{\alpha\beta}\epsilon_{AB} \psi^{\beta B}
\ee
The 2-point functions are
\be
<\psi^{\alpha A}(z)\psi^\dagger_{\beta B}(w)>=\delta^\alpha_\beta\delta^A_B{1\over z-w}, ~~~
<\psi^{\alpha A}(z)\psi^{\beta B}(w)>=-\epsilon^{\alpha\beta}\epsilon^{AB}{1\over z-w}
\ee
where we have defined the $\epsilon$ symbol as
\be
\epsilon_{12}=1, ~~~\epsilon^{12}=-1, ~~~
\psi_A=\epsilon_{AB}\psi^B, ~~~
\psi^A=\epsilon^{AB}\psi_B
\ee
There are 4 real left moving bosons $X_1, X_2, X_3, X_4$ which can be grouped into a matrix 
\be
X_{A\dot A}= \sqi X_i \sigma_i=\sqi\pmatrix { X_3+iX_4& X_1-iX_2\cr X_1+iX_2&-X_3+iX_4\cr}
\ee
where $\sigma_i=\sigma_a, iI$. This gives
\be
(X^*)^{A\dot A}=\sqi\pmatrix { X_3-iX_4& X_1+iX_2\cr X_1-iX_2&-X_3-iX_4\cr}
\ee
and the reality condition
\be
(X_{A\dot A})^*=X^{A\dot A}
\ee
The 2-point functions are
\be
<\p X_{A\dot A}(z) (\p X^\dagger)^{B\dot B}(w)>=-{1\over (z-w)^2}\delta^B_A\delta^{\dot B}_{\dot A}, ~~~
<\p X_{A\dot A}(z) \p X_{B\dot B}(w)>={1\over (z-w)^2}\epsilon_{AB}\epsilon_{\dot A\dot B}
\ee

The chiral algebra is generated by the operators
\be
J^a=-{1\over 4}(\psi^\dagger)_{\alpha A} (\sigma^{Ta})^\alpha{}_\beta \psi^{\beta A}
\ee
\be
G^\alpha_{\dot A}= \psi^{\alpha A} \p X_{A\dot A}, ~~~(G^\dagger)_{\alpha}^{\dot A}=(\psi^\dagger)_{\alpha A} \p (X^\dagger)^{A\dot A}
\ee
\be
T=-{1\over 2} (\p X^\dagger)^{A\dot A}\p X_{A\dot A}-{1\over 2} (\psi^\dagger)_{\alpha A} \p \psi^{\alpha A}
\ee
\be
(G^\dagger)_{\alpha}^{\dot A}=-\epsilon_{\alpha\beta} \epsilon^{\dot A\dot B}G^\beta_{\dot B}, ~~~~G^{\alpha}_{\dot A}=-\epsilon^{\alpha\beta} \epsilon_{\dot A\dot B}(G^\dagger)_\beta^{\dot B}
\ee
These operators generate the algebra
\be
J^a(z) J^b(w)\sim \delta^{ab} {\h\over (z-w)^2}+i\epsilon^{abc} {J^c\over z-w}
\ee
\be
J^a(z) G^\alpha_{\dot A} (z')\sim {1\over (z-z')}\h (\sigma^{aT})^\alpha{}_\beta G^\beta_{\dot A}
\ee
\be
G^\alpha_{\dot A}(z) (G^\dagger)^{\dot B}_\beta(z')\sim -{2\over (z-z')^3}\delta^\alpha_\beta \delta^{\dot B}_{\dot A}- \delta^{\dot B}_{\dot A}  (\sigma^{Ta})^\alpha{}_\beta [{2J^a\over (z-z')^2}+{\p J^a\over (z-z')}]
-{1\over (z-w)}\delta^\alpha_\beta \delta^{\dot B}_{\dot A}T
\ee
\be
T(z)T(z')\sim {3\over (z-z')^4}+{2T\over (z-z')^2}+{\p T\over (z-z')}
\ee
\be
T(z) J^a(z')\sim {J^a\over (z-z')^2}+{\p J^a\over (z-z')}
\ee
\be
T(z) G^\alpha_{\dot A}\sim {{3\over 2}G^\alpha_{\dot A}\over (z-z')^2}  + {\p G^\alpha_{\dot A}\over (z-z')} 
\ee

Note that
\be
J^a(z) \psi^{\gamma C}(w)\sim {1\over 2} {1\over z-w} (\sigma^{aT})^\gamma{}_\beta \psi^{\beta C}
\ee

The above OPE algebra gives the commutation relations
\begin{eqnarray}
\com{J^a_m}{J^b_n} &=& \frac{m}{2}\delta^{ab}\delta_{m+n,0} + i{\epsilon^{ab}}_c J^c_{m+n}
            \\
\com{J^a_m}{G^\alpha_{\dot{A},n}} &=& \frac{1}{2}{(\sigma^{aT})^\alpha}_\beta G^\beta_{\dot{A},m+n}
             \\
\ac{G^\alpha_{\dot{A},m}}{G^\beta_{\dot{B},n}} &=& \hspace*{-4pt}\epsilon_{\dot{A}\dot{B}}\bigg[
   (m^2 - \frac{1}{4})\epsilon^{\alpha\beta}\delta_{m+n,0}
  + (m-n){(\sigma^{aT})^\alpha}_\gamma\epsilon^{\gamma\beta}J^a_{m+n}
  + \epsilon^{\alpha\beta} L_{m+n}\bigg]\\
\com{L_m}{L_n} &=& \frac{m(m^2-\frac{1}{4})}{2}\delta_{m+n,0} + (m-n)L_{m+n}\\
\com{L_m}{J^a_n} &=& -n J^a_{m+n}\\
\com{L_m}{G^\alpha_{\dot{A},n}} &=& \left(\frac{m}{2}-n\right)G^\alpha_{\dot{A},m+n}
\end{eqnarray}

\section{Showing $G^-\sigma_2^+\sim G^+\sigma_2^-$} \label{ap:EqOfTwist}

Here we prove that the two  operators
\be
{1\over 2\pi i} \int _{w_0} dw G^-_{\dot A} (w) \sigma_2^+(w_0)
\label{apbone}
\ee
and
\be
{1\over 2\pi i} \int _{w_0} dw G^+_{\dot A} (w) \sigma_2^-(w_0)
\label{apbtwo}
\ee 
are proportional to each other, so we do not need to add over both
these possibilities in the deformation operator (\ref{deformation}).

Under the map $z=e^w$ (\ref{apbone}) gives
\be
{1\over 2\pi i} \int _{w_0} dw G^-_{\dot A} (w) \sigma_2^+(w_0)={1\over 2\pi i} \int _{z_0} dz G^-_{\dot A} (z) z^{\h}\sigma_2^+(z_0)
\ee
Under the map $z=z_0+t^2$ we get
\be
 {1\over 2\pi i} \int _{z_0} dz G^-_{\dot A} (z) z^{\h}\sigma_2^+(z_0)={1\over 2\pi i} \int_{t=0} dt G^-_{\dot A}(t) (2t)^{-\h} (z_0+t^2)^\h |0^+_R\rangle_t
 \ee
 where we have noted that the twist $\sigma_2^+$ maps to the positive charge Ramond vacuum in the $t$ plane
 \be
 \sigma_2^+(z_0)\r |0^+_R\rangle_t
 \ee
(The possible constant in this relation is fixed to unity  by the definitions in section \ref{normsection}.)
We perform  a spectral flow in this $t$ plane with spectral flow parameter $\alpha=-1$. Noting that under this spectral flow we have $G^-_{\dot A}(t)\r t^{-\h} G^-_{\dot A}(t)$ we get
\bea
 {1\over 2\pi i} \int_{t=0} dt G^-_{\dot A}(t) (2t)^{-\h} (z_0+t^2)^\h |0^+_R\rangle_t &\r&
 {1\over 2\pi i} \int_{t=0} dt G^-_{\dot A}(t) (2t)^{-\h} (z_0+t^2)^\h t^{-\h}|0\rangle_t\nonumber \\ 
 &=&2^{-\h} z_0^\h  \t G^-_{\dot A, -{3\over 2}}|0\rangle_t
 \eea
 where the tilde symbol over the supercurrent denotes operators in the $t$ plane
  \be
\t G^-_{\dot A, -{3\over 2}}={1\over 2\pi i} \int_{t=0} dt t^{-1}G^-_{\dot A} (t) 
\label{apbfour}
 \ee
  
Now we perform similar manipulations on (\ref{apbtwo}). The map to the $t$ plane gives
\be
 {1\over 2\pi i} \int _{z_0} dz G^+_{\dot A} (z) z^{-\h}\sigma_2^-(z_0)={1\over 2\pi i} \int_{t=0} dt G^+_{\dot A}(t) (2t)^{-\h} (z_0+t^2)^\h |0^-_R\rangle_t
 \ee
Under the  spectral flow with parameter $\alpha=-1$ the state $|0^-_R\rangle_t$ goes to a state with dimension $1$ and charge $-1$. There is just one state with these quantum numbers, so we can write
\be
|0^-_R\rangle_t\r C \t J^-_{-1}|0\rangle_t
\ee
Noting that under this spectral flow we have $G^+_{\dot A}(t)\r t^\h G^+_{\dot A}(t)$ we get
after the spectral flow
\bea
{1\over 2\pi i} \int_{t=0} dt G^+_{\dot A}(t) (2t)^{-\h} (z_0+t^2)^\h |0^-_R\rangle_t
&\r& {1\over 2\pi i} \int_{t=0} dt G^+_{\dot A}(t) (2t)^{-\h} (z_0+t^2)^\h t^{\h}C \t J^-_{-1}|0\rangle_t\nonumber \\
 &=&2^{-\h} z_0^\h C \t G^+_{\dot A, -{1\over 2}}\t J^-_{-1}|0\rangle_t
 \eea
We can now use the commutation relation
\be
[\t G^+_{\dot A, -\h}, \t J^-_{-1}]=-\t G^+_{\dot A, -{3\over 2}}
\ee
together with the fact that $G^+_{\dot A, -\h}|0\rangle_t=0$ to write
\be
2^{-\h} z_0^\h C \t G^+_{\dot A, -{1\over 2}}\t J^-_{-1}|0\rangle_t=-2^{-\h} z_0^\h C \t G^+_{\dot A, -{3\over 2}}|0\rangle_t
\ee
which is proportional to (\ref{apbfour}).

\section{The exponential ansatz }\label{cc}

In this appendix we check that the exponential ansatz (\ref{pfive}) for the bosonic excitations is indeed correct; the fermionic case works in a similar way.

The bosonic fields completely decouple from the fermionic fields. Further, the set $\{\alpha_{++, m}, \alpha_{--, n}\}$ decouples from the set
$\{ \alpha_{+-,m},\alpha_{-+, n}\}$. Thus in what follows we write only the modes $\{\alpha_{++, m}, \alpha_{--, n}\}$. 

We have from the basic relation (\ref{master})
\be
\langle 0_{R,-}|\Big (\alpha_{++, n_1}\alpha_{--, n_2}\dots \Big )\sigma^+_2(w_0)|0^-_R\rangle^{(1)}\otimes |0^-_R\rangle^{(2)}={}_t\langle 0 |\Big (\alpha'_{++, n_1}\alpha'_{--, n_2}\dots \Big )|0\rangle_t
\label{apptwo}
\ee
where the operators $\alpha'$ arise from following the various coordinate changes and spectral flows that bring us from the original operators $\alpha$ on the cylinder to operators on the $t$ plane (with the state $|0\rangle_t$ at $t=0$). 
If we can understand all amplitudes of this type, then we will have a complete understanding of the state $\sigma_2^+(w_0)|0^-_R\rangle^{(1)}|0^-_R\rangle^{(2)}$. Our ansatz for this state is
\be
\sigma_2^+(w_0)|0^-_R\rangle^{(1)}|0^-_R\rangle^{(2)}=e^{-\sum_{m>0, n>0}\gamma^B_{mn}\alpha_{++,m}\alpha_{--,n}}|0^-_R\rangle
\label{appfive}
\ee

The initial operators $\alpha$ are given by eq.(\ref{qaone}), and their map to the final form in the $t$ plane is given by eq.(\ref{qatwo}). We can expand the latter form in terms of natural modes on the $t$ plane
(\ref{qathree})
\bea
\alpha_{A\dot A, n}&\r& {1\over 2\pi} \int_{t=\infty} \p_t X_{A\dot A}(t) (z_0+t^2)^{{n\over 2}} dt\nonumber \\
&=&{1\over 2\pi} \int_{t=\infty} \p_t dt X_{A\dot A}(t)\sum_{k\ge 0} {}^{n\over 2} C_k z_0^k t^{n-2k} \nonumber \\
&=&\sum_{k\ge 0} {}^{n\over 2} C_kz_0^k \t\alpha_{A\dot A, n-2k}\label{appsix}
\eea
All we need to know is that this is a linear relation 
\be
\alpha_{A\dot A, n}=\sum_{p=-\infty}^\infty B_{n,p}\t\alpha_{A\dot A, p}
\label{appthree}
\ee
with some constant coefficients $B_{np}$. Since the relation (\ref{appthree}) is linear in the field operators, we will have as many operators inserted between the brackets $ (  )$ on the RHS of (\ref{apptwo}) as on the LHS. But on the RHS we  just have these mode operators sandwiched between the $t$ plane vacuum state. Thus the amplitude will be evaluated by Wick contractions between these operators. From this fact we can immediately note two things: we must have an even number of insertions, and there must be an equal number of $\alpha_{++}$ and $\alpha_{--}$ modes. 

Let us start with the simplest case: two operator insertions, which is computation we encountered in finding $\gamma^B_{mn}$. We have (with $n_1>0, n_2>0$)
\be
 \langle 0_{R,-}|\Big (\alpha_{++, n_1}\alpha_{--, n_2}\Big )\sigma^+_2(w_0)|0^-_R\rangle^{(1)}\otimes |0^-_R\rangle^{(2)}={}_t\langle 0 |\Big (\alpha'_{++, n_1}\alpha'_{--, n_2}\Big )|0\rangle_t
 \label{appone}
\ee
With the ansatz (\ref{appfive}), the LHS gives
\begin{eqnarray}
&&\hspace*{-70pt}\langle 0_{R,-}|
\Big(\alpha_{++, n_1}\alpha_{--, n_2}\Big)
\sigma^+_2(w_0)|0^-_R\rangle^{(1)}\otimes |0^-_R\rangle^{(2)}\nonumber\\
&&\hspace*{-55pt}=\langle 0_{R,-}|
\Big(\alpha_{++, n_1}\alpha_{--, n_2}\Big)
e^{-\sum_{m_1>0, m_2>0}\gamma^B_{m_1m_2}\alpha_{++,-m_1}\alpha_{--,-m_2}}|0^-_R\rangle
\end{eqnarray}
The contribution to this amplitude comes from expanding the exponential to first order
giving
\be
\langle 0_{R,-}|
\Big(\alpha_{++, n_1}\alpha_{--, n_2}\Big)(-1)\hspace{-10pt}
\sum_{m_1>0, m_2>0}\hspace{-10pt}\gamma^B_{m_1m_2}\alpha_{++,-m_1}\alpha_{--,-m_2}|0^-_R\rangle
=(-1) n_1n_2 \gamma^B_{n_2n_1}
\ee
The RHS of (\ref{appone}) gives
\be
{}_t\langle 0 |\sum_{p_1, p_2} B_{n_1,p_1}B_{n_2,p_2}
  \t \alpha_{++,p_1}\t \alpha_{--,p_2}|0\rangle_t=\sum_{p_1>0} p_1 B_{n_1,p_1}B_{n_2, -p_1}
\ee
Thus we get the relation
\be
(-1)n_1n_2 \gamma^B_{n_2n_1}=\sum_{p_1>0} p_1 B_{n_1,p_1}B_{n_2, -p_1}
\label{appseven}
\ee
which gives the $\gamma^B_{mn}$ in (\ref{gammaB}) when we use (\ref{appsix}).

Let us now consider the next simplest case: four operator insertions. The LHS of (\ref{apptwo}) gives ($n_1, n_2, n_3, n_4>0$)
\be
\langle 0_{R,-}|\Big (\alpha_{++, n_1}\alpha_{--, n_2}\alpha_{++, n_3}\alpha_{--, n_4}\Big )\sigma^+_2(w_0)|0^-_R\rangle^{(1)}\otimes |0^-_R\rangle^{(2)}
\ee
We must now expand the exponential in the ansatz (\ref{appfive}) to second order, 
getting for the LHS of (\ref{apptwo})
\begin{eqnarray}
&&\hspace*{-60pt}{}^{(2)}\langle 0_{R,-}|
\Big(\alpha_{++, n_1}\alpha_{--, n_2}\alpha_{++, n_3}\alpha_{--, n_4}\Big) 
{1\over 2!}(-1)^2\nonumber\\
&&\hspace*{-30pt}\times
\Big(\sum_{m_1>0, m_2>0}\gamma^B_{m_1m_2}\alpha_{++,-m_1}\alpha_{--,-m_2}\Big)
\Big(\sum_{m_3>0, m_4>0}\gamma^B_{m_3m_4}\alpha_{++,-m_3}\alpha_{--,-m_4}\Big)|0^-_R\rangle
\end{eqnarray}
This gives
\be
{1\over 2!}(-1)^2(2!)n_1n_2n_3n_4\Big ( \gamma^B_{n_2n_1}\gamma^B_{n_4n_3} +  \gamma^B_{n_4n_1}\gamma^B_{n_2n_3} \Big )
\label{appeight}
\ee
where the factor $(2!)$ comes from the fact that the set $\alpha_{++, n_1}\alpha_{--,n_2}$ can contract with the operators from either of the two $\gamma^B$ factors. 

The RHS of (\ref{apptwo}) gives
\begin{eqnarray}
&{}_t\langle 0 |&\sum_{p_1, p_2,p_3,p_4} B_{n_1,p_1}B_{n_2,p_2}B_{n_3,p_3}B_{n_4,p_4}
 \t \alpha_{++,p_1}\t \alpha_{--,p_2}\t\alpha_{++,p_3}\t\alpha_{--,p_4}|0\rangle_t\nonumber\\
&&= \Big(\sum_{p_1>0} p_1 B_{n_1,p_1}B_{n_2, -p_1}\Big)
 \Big(\sum_{p_3>0} p_3 B_{n_3,p_3}B_{n_4, -p_3}\Big)\nonumber\\
&&\quad+\Big(\sum_{p_1>0} p_1 B_{n_1,p_1}B_{n_4, -p_1}\Big)
 \Big(\sum_{p_3>0} p_3 B_{n_3,p_3}B_{n_2, -p_3}\Big)
\end{eqnarray}
On using (\ref{appseven}) this gives
\be
n_1n_2n_3n_4\Big ( \gamma^B_{n_2n_1}\gamma^B_{n_4n_3} +  \gamma^B_{n_4n_1}\gamma^B_{n_2n_3} \Big )
\ee
which agrees with (\ref{appeight}).

Thus we have verified the ansatz to order four in the bosonic field operators. Proceeding in this way we can verify the complete exponential ansatz.

The fermionic case is similar. Modes on the cylinder map linearly to the modes for the case where we are on the $t$ plane and we have the NS vacuum at $t=0$. On this $t$ plane the modes must appear in pairs to allow the amplitude to be nonvanishing, thus we must have an even number of modes in each term in the ansatz. Our ansatz allows all modes that are nonvanishing on the chosen vacuum state; thus the situation is similar to the bosonic case where we allowed all negative index bosonic operators in the ansatz. Thus the fermionic part of the ansatz can be verified in the same way as the bosonic part.


\begin{thebibliography}{10}

\bibitem{counting}
A.~Sen,
  Nucl.\ Phys.\  B {\bf 440}, 421 (1995)
  [arXiv:hep-th/9411187].
A.~Sen,
  Mod.\ Phys.\ Lett.\  A {\bf 10}, 2081 (1995)
  [arXiv:hep-th/9504147].
A.~Strominger and C.~Vafa,
  Phys.\ Lett.\  B {\bf 379}, 99 (1996)
  [arXiv:hep-th/9601029].
A.~Dabholkar,
  Phys.\ Rev.\ Lett.\  {\bf 94}, 241301 (2005)
  [arXiv:hep-th/0409148].
A.~Dabholkar, R.~Kallosh and A.~Maloney,
  JHEP {\bf 0412}, 059 (2004)
  [arXiv:hep-th/0410076].


\bibitem{radiation}
C.~G.~Callan and J.~M.~Maldacena,
  Nucl.\ Phys.\  B {\bf 472}, 591 (1996)
  [arXiv:hep-th/9602043];
A.~Dhar, G.~Mandal and S.~R.~Wadia,
  Phys.\ Lett.\  B {\bf 388}, 51 (1996)
  [arXiv:hep-th/9605234];
S.~R.~Das and S.~D.~Mathur,
  Nucl.\ Phys.\ B {\bf 478}, 561 (1996)
  [arXiv:hep-th/9606185];
S.~R.~Das and S.~D.~Mathur,
  Nucl.\ Phys.\  B {\bf 482}, 153 (1996)
  [arXiv:hep-th/9607149];
J.~M.~Maldacena and A.~Strominger,
  Phys.\ Rev.\  D {\bf 55}, 861 (1997)
  [arXiv:hep-th/9609026].
S.~D.~Mathur,
  Nucl.\ Phys.\  B {\bf 514}, 204 (1998)
  [arXiv:hep-th/9704156];
S.~S.~Gubser,
  Phys.\ Rev.\  D {\bf 56}, 4984 (1997)
  [arXiv:hep-th/9704195].
O.~J.~C.~Dias, R.~Emparan and A.~Maccarrone,
  ``Microscopic Theory of Black Hole Superradiance,''
  Phys.\ Rev.\  D {\bf 77}, 064018 (2008)
  [arXiv:0712.0791 [hep-th]].
B.~D.~Chowdhury and A.~Virmani,
  arXiv:1001.1444 [hep-th].


\bibitem{fuzzballs}
O.~Lunin and S.~D.~Mathur,
  Nucl.\ Phys.\  B {\bf 615}, 285 (2001)
  [arXiv:hep-th/0107113].
O.~Lunin and S.~D.~Mathur,
  Nucl.\ Phys.\  B {\bf 623}, 342 (2002)
  [arXiv:hep-th/0109154].
O.~Lunin and S.~D.~Mathur,
  Phys.\ Rev.\ Lett.\  {\bf 88}, 211303 (2002)
  [arXiv:hep-th/0202072].
O.~Lunin, J.~Maldacena and L.~Maoz,
  [arXiv:hep-th/0212210].
S.~D.~Mathur, A.~Saxena and Y.~K.~Srivastava,
  Nucl.\ Phys.\ B {\bf 680}, 415 (2004)
  [arXiv:hep-th/0311092].
S.~Giusto, S.~D.~Mathur and A.~Saxena,
  Nucl.\ Phys.\ B {\bf 701}, 357 (2004)
  [arXiv:hep-th/0405017].
O.~Lunin,
  JHEP {\bf 0404}, 054 (2004)
  [arXiv:hep-th/0404006].
S.~Giusto, S.~D.~Mathur and A.~Saxena,
  Nucl.\ Phys.\  B {\bf 710}, 425 (2005)
  [arXiv:hep-th/0406103].
I.~Bena and N.~P.~Warner,
  Adv.\ Theor.\ Math.\ Phys.\  {\bf 9}, 667 (2005)
  [arXiv:hep-th/0408106];
S.~D.~Mathur,
  Fortsch.\ Phys.\  {\bf 53}, 793 (2005)
  [arXiv:hep-th/0502050];
I.~Bena and N.~P.~Warner,
  Phys.\ Rev.\  D {\bf 74}, 066001 (2006)
  [arXiv:hep-th/0505166];
P.~Berglund, E.~G.~Gimon and T.~S.~Levi,
  JHEP {\bf 0606}, 007 (2006)
  [arXiv:hep-th/0505167];
M.~Taylor,
  JHEP {\bf 0603}, 009 (2006)
  [arXiv:hep-th/0507223];
A.~Saxena, G.~Potvin, S.~Giusto and A.~W.~Peet,
  JHEP {\bf 0604}, 010 (2006)
  [arXiv:hep-th/0509214];
S.~D.~Mathur,
  Class.\ Quant.\ Grav.\  {\bf 23}, R115 (2006)
  [arXiv:hep-th/0510180];
I.~Bena, C.~W.~Wang and N.~P.~Warner,
  Phys.\ Rev.\  D {\bf 75}, 124026 (2007)
  [arXiv:hep-th/0604110];
V.~Balasubramanian, E.~G.~Gimon and T.~S.~Levi,
  JHEP {\bf 0801}, 056 (2008)
  [arXiv:hep-th/0606118];
I.~Bena, C.~W.~Wang and N.~P.~Warner,
  JHEP {\bf 0611}, 042 (2006)
  [arXiv:hep-th/0608217];
K.~Skenderis and M.~Taylor,
  Phys.\ Rev.\ Lett.\  {\bf 98}, 071601 (2007)
  [arXiv:hep-th/0609154].
J.~Ford, S.~Giusto and A.~Saxena,
  Nucl.\ Phys.\  B {\bf 790}, 258 (2008)
  [arXiv:hep-th/0612227];
I.~Bena and N.~P.~Warner,
  arXiv:hep-th/0701216;
I.~Kanitscheider, K.~Skenderis and M.~Taylor,
  JHEP {\bf 0706}, 056 (2007)
  [arXiv:0704.0690 [hep-th]];
I.~Bena, N.~Bobev and N.~P.~Warner,
  JHEP {\bf 0708}, 004 (2007)
  [arXiv:0705.3641 [hep-th]];
S.~Giusto and A.~Saxena,
  Class.\ Quant.\ Grav.\  {\bf 24}, 4269 (2007)
  [arXiv:0705.4484 [hep-th]];
E.~G.~Gimon and T.~S.~Levi,
  JHEP {\bf 0804}, 098 (2008)
  [arXiv:0706.3394 [hep-th]].
I.~Bena, C.~W.~Wang and N.~P.~Warner,
  JHEP {\bf 0807}, 019 (2008)
  [arXiv:0706.3786 [hep-th]].
J.~Ford, S.~Giusto, A.~Peet and A.~Saxena,
  Class.\ Quant.\ Grav.\  {\bf 25}, 075014 (2008)
  [arXiv:0708.3823 [hep-th]];
S.~Giusto, S.~F.~Ross and A.~Saxena,
  JHEP {\bf 0712}, 065 (2007)
  [arXiv:0708.3845 [hep-th]].
K.~Skenderis and M.~Taylor,
  Phys.\ Rept.\  {\bf 467}, 117 (2008)
  [arXiv:0804.0552 [hep-th]];
   V.~Jejjala, O.~Madden, S.~F.~Ross and G.~Titchener,
  Phys.\ Rev.\  D {\bf 71}, 124030 (2005)
  [arXiv:hep-th/0504181];
J.~H.~Al-Alawi and S.~F.~Ross,
  JHEP {\bf 0910}, 082 (2009)
  [arXiv:0908.0417 [hep-th]];
 I.~Bena, S.~Giusto, C.~Ruef and N.~P.~Warner,
  JHEP {\bf 0911}, 089 (2009)
  [arXiv:0909.2559 [hep-th]];
N.~Bobev and C.~Ruef,
  JHEP {\bf 1001}, 124 (2010)
  [arXiv:0912.0010 [hep-th]].



\bibitem{adscft}
  J.~M.~Maldacena,
  Adv.\ Theor.\ Math.\ Phys.\  {\bf 2}, 231 (1998)
  [Int.\ J.\ Theor.\ Phys.\  {\bf 38}, 1113 (1999)]
  [arXiv:hep-th/9711200];
  E.~Witten,
  Adv.\ Theor.\ Math.\ Phys.\  {\bf 2}, 253 (1998)
  [arXiv:hep-th/9802150];
  S.~S.~Gubser, I.~R.~Klebanov and A.~M.~Polyakov,
  Phys.\ Lett.\  B {\bf 428}, 105 (1998)
  [arXiv:hep-th/9802109].


\bibitem{orbifold}
G.~E.~Arutyunov and S.~A.~Frolov,
  Theor.\ Math.\ Phys.\  {\bf 114}, 43 (1998)
  [arXiv:hep-th/9708129];
G.~E.~Arutyunov and S.~A.~Frolov,
  Nucl.\ Phys.\  B {\bf 524}, 159 (1998)
  [arXiv:hep-th/9712061];
J.~de Boer,
  Nucl.\ Phys.\  B {\bf 548}, 139 (1999)
  [arXiv:hep-th/9806104];
N.~Seiberg and E.~Witten,
  JHEP {\bf 9904}, 017 (1999)
  [arXiv:hep-th/9903224];
R.~Dijkgraaf,
  Nucl.\ Phys.\  B {\bf 543}, 545 (1999)
  [arXiv:hep-th/9810210];
F.~Larsen and E.~J.~Martinec,
  JHEP {\bf 9906}, 019 (1999)
  [arXiv:hep-th/9905064];
A.~Jevicki, M.~Mihailescu and S.~Ramgoolam,
  Nucl.\ Phys.\  B {\bf 577}, 47 (2000)
  [arXiv:hep-th/9907144];
  A.~Pakman, L.~Rastelli and S.~S.~Razamat,
  JHEP {\bf 0910}, 034 (2009)
  [arXiv:0905.3448 [hep-th]];
A.~Pakman, L.~Rastelli and S.~S.~Razamat,
  Phys.\ Rev.\  D {\bf 80}, 086009 (2009)
  [arXiv:0905.3451 [hep-th]];
A.~Pakman, L.~Rastelli and S.~S.~Razamat,
  arXiv:0912.0959 [hep-th].



\bibitem{deformation}
  J.~R.~David, G.~Mandal and S.~R.~Wadia,
  Nucl.\ Phys.\  B {\bf 564}, 103 (2000)
  [arXiv:hep-th/9907075];
   J.~Gomis, L.~Motl and A.~Strominger,
  JHEP {\bf 0211}, 016 (2002)
  [arXiv:hep-th/0206166];
E.~Gava and K.~S.~Narain,
  JHEP {\bf 0212}, 023 (2002)
  [arXiv:hep-th/0208081].



\bibitem{spectral}
  A.~Schwimmer and N.~Seiberg,
  Phys.\ Lett.\  B {\bf 184}, 191 (1987).

\bibitem{lm2}
  O.~Lunin and S.~D.~Mathur,
  Commun.\ Math.\ Phys.\  {\bf 227}, 385 (2002)
  [arXiv:hep-th/0103169].

\bibitem{acm} 
  S.~G.~Avery, B.~D.~Chowdhury and S.~D.~Mathur,
  arXiv:0906.2015 [hep-th].

\bibitem{lm1}
  O.~Lunin and S.~D.~Mathur,
  Commun.\ Math.\ Phys.\  {\bf 219}, 399 (2001)
  [arXiv:hep-th/0006196].



\bibitem{microrad}
V.~Cardoso, O.~J.~C.~Dias and R.~C.~Myers,
  Phys.\ Rev.\  D {\bf 76}, 105015 (2007)
  [arXiv:0707.3406 [hep-th]].
  B.~D.~Chowdhury and S.~D.~Mathur,
  Class.\ Quant.\ Grav.\  {\bf 25}, 135005 (2008)
  [arXiv:0711.4817 [hep-th]].
  B.~D.~Chowdhury and S.~D.~Mathur,
  arXiv:0806.2309 [hep-th].
  B.~D.~Chowdhury and S.~D.~Mathur,
  Class.\ Quant.\ Grav.\  {\bf 26}, 035006 (2009)
  [arXiv:0810.2951 [hep-th]].
  S.~G.~Avery and B.~D.~Chowdhury,
  arXiv:0907.1663 [hep-th].

\end{thebibliography}
\end{document}